\documentclass[letterpaper, 10 pt, conference]{ieeeconf}  
\usepackage[utf8]{inputenc}

\IEEEoverridecommandlockouts                              
\usepackage{hyperref}

\usepackage{graphics} 
\usepackage{amsmath} 
\usepackage{amssymb}  
\usepackage{blkarray}
\usepackage{tikz}
\usepackage{xcolor}
\usepackage{pgfplots}
\pgfplotsset{compat=newest}
\usepgfplotslibrary{groupplots}
\usepgfplotslibrary{dateplot}
\usepackage{graphicx}
\usetikzlibrary{shadows.blur,positioning,calc,arrows.meta}
\usepackage{algpseudocode}
\usepackage{algorithm}
\usepackage{biblatex}
\title{\LARGE \bf
Reduced-Order Modeling of Thermal Dynamics in District Energy Networks using Spectral Clustering
}

\author{Johan Simonsson$^{1}{}^{2}{}^\dagger$, Khalid Tourkey Atta$^{1}$, Wolfgang Birk$^{1}$
\thanks{This work was financially supported by the Swedish Energy Agency under Grant 43090-2, Cloudberry Datacenters, and ERA-Net Smart Energy Systems through the project FlexiSync, funded by the European Union's Horizon 2020 research and innovation program under Grant Agreement No. 775970 (RegSys).}
\thanks{$^{1}$Johan Simonsson, Khalid Tourkey Atta and Wolfgang Birk are with Luleå University of Technology
$^{2}$ Johan Simonsson is with Optimation AB, Uppsala, Sweden
$^{\dagger}$ Corresponding author, {\tt\small johan.simonsson@ltu.se}}%
}

\begin{document}

\maketitle
\thispagestyle{empty}
\pagestyle{empty}

\begin{abstract}
Simulation of thermal dynamics in city-scale district energy grids often becomes computationally prohibitive for long simulation runs. Current model order reduction methods offer limited interpretability with regards to the non-reduced system, and are not in general applicable for e.g., varying flow rates, multiple producers, or changing flow directions. This article presents a novel method based on graph theory that approximates the solution of an optimization problem that minimizes the local truncation error for heat transport in the grid. It is shown that the method can be used to reduce the thermal dynamic model of a city-scale energy grid, resulting in a coarser temporal and spatial resolution. The relative root mean square error was 2.3\% for the temperature in the evaluation scenario, comparing the reduced-order system with the non-reduced system at the instances of the coarser time-step. 
\end{abstract}


\section{INTRODUCTION}
Simulation of district energy networks is a popular subject of scientific articles, where simulations are used e.g. for the planning of new district energy grids, optimization of production planning and thermal storage \cite{Vandermeulen2018}, and to explore various scenarios such as lowered supply temperatures and including waste heat from data centers. 

There are several domain-specific commercial tools available for simulation of district energy networks, where most are focused on the steady-state analysis of pressure, flow, and temperatures \cite{schweiger_district_2018}. Dynamic simulation of the temperature in district energy grids on a city scale generally suffers from limited practical usability due to the computational demands of dynamic simulation \cite{simonsson_experiences_2021}. The computational performance is highly dependent on the scale of the problem and the resolution of interest -- both the spatial and temporal resolution. The scale of the problem is usually given beforehand, but the resolution of interest is dependent on the use cases, and can also vary within a specific use case. It would thus be ideal to find methods that can adapt the spatial and time resolution of the simulation between simulation runs, or even dynamically during simulation.

Existing methods to reduce the computational demand can be roughly split into three categories. The first category consists of methods to simplify the model by merging nodes and pipes in a structured manner based on the physics of district energy grids. The most popular methods are called the Danish and the German method, respectively \cite{falay_enabling_2020}. However, both methods are limited in that they do not preserve the original structure of the grid, and thus the states are not directly interpretable with regards to the original system. Moreover, the methods cannot handle changing flow directions or multiple production units. A second category consists of model order reduction techniques for dynamic systems on state space form. Methods include Proper Orthogonal Decomposition (POD) and Hankel Model Order Reduction \cite{antoulas_approximation_2005}. Most of these methods rely on the system being linear and time-invariant, which is not true for the thermal dynamics of pipe flow with varying flow rates. Moreover, the POD method aims to find the most influential states, rather than averaging over (aggregating) many states. A third, somewhat related, category consists of methods for Computation Fluid Dynamics (CFD), e.g. various multi-grid methods \cite{trottenberg_multigrid_2000}. These methods are mostly focused on higher dimensions or higher accuracy than in this article, and are generally not applicable.

The article is structured as follows. First, the mathematical background for the differential equations of heat transport in a district energy pipe is presented, followed by a short introduction to the graph theory used in the article. Next, the two are combined to show how advection on a graph can be modeled. General reduced order models of a linear time-varying state space system for advection on a graph are then presented. 

The article then proposes a novel method based on graph theory and the partial differential equations of advection with corresponding discretization methods, posing this as an optimization problem where the spatial resolution of the grid is optimized with regards to the time resolution of interest, minimizing the local truncation error. To avoid falling into the NP-hardness trap that is common for algorithms on graphs, the solution to the optimization problem is approximated by solving a generalized eigenvalue problem using the so-called graph Laplacian(s).  While there are generic algorithms for clustering (sometimes called sparsening or coarsening of graphs) \cite{loukas_graph_2019}, these cannot be directly applied to a district energy grid without considering the underlying physics. 

In the results section, the performance of the clustering is evaluated with regards to a non-reduced city-scale grid, and it is shown that the relative root mean square (rRMSE) error is $2.3\%$ for the temperature of all the nodes of the original model when comparing $5s$ time steps with $600s$ time steps at the points of the coarser time step. The computational performance is improved with a speedup factor of $\approx 90$ for the evaluation case. 

The scientific contribution of the article is showing how a reduced-order model of a district energy grid can be constructed, showing how finding the reduced-order model can be formulated as an optimization problem that minimizes the local truncation error of the discretization with regards to the chosen time resolution, and proposing a computationally efficient method to approximately solve the problem using spectral graph theory.

\section{BACKGROUND}
\subsection{Thermal dynamics of pipe flow}
For the scope of the article, the problem is limited to a temporal resolution ranging from seconds to hours, so that pressure dynamics can be modeled as static, and the water can be assumed incompressible. Further, it is assumed that the flow rate of each pipe is known. For a complete simulation model, the pressure-flow dynamics need to be modeled, e.g., interfacing a separate model or using iterative methods \cite{todini_unified_2013}. 

The thermal dynamics of a district heating pipe can be described by a 1D partial differential equation (PDE), where axial diffusion, pressure losses, dissipation and wall friction have been shown to have a negligible impact on the temperature for the operational ranges of district heating, and are thus neglected \cite{van_der_heijde_dynamic_2017}. The remaining PDE is 
\begin{align}
    \label{eq:pipe_pde}
    \rho c_p A \frac{dx(z,t)}{dt} + q(z, t) = -\rho c_p A v(t)  \frac{dx(z,t)}{dz},
\end{align}
\noindent where $x$ is the temperature, $t$ time, and $z$ the axial direction. $\rho$ is the density, $c_p$ is the heat capacity, and $A$ is the cross sectional area of the pipe, where all three are modeled as constant. $q(z, t)$ is the heat loss to the surroundings and $v(t)$ the flow velocity in the axial direction. For the remainder of the paper, the heat losses to surroundings are left out of the analysis for notational convenience, but are straight forward to include.

To solve the PDE in Equation \eqref{eq:pipe_pde} numerically the pipe is discretized along the pipe length using a finite volume method and an upwind discretization scheme \cite{sartor_comparative_2018}. Using subscripts for spatial steps, the balance equation for each finite volume of the pipe can be written as
\begin{align}
    \label{eq:discretized_pipe}
    \Dot{x}_i(t) = \frac{v_i(t)}{\Delta z}(x_i(t) - x_{i-1}(t)).
\end{align}

A necessary condition for stability of an explicit time integration schemes is that the volume of flow through a volume during a time step of the solver does not exceed the size of the volume. This is known as the Courant-Friedrichs-Lewy (CFL) condition and can be written as 
\begin{align}
    \label{eq:CFL}
    \frac{v_{sup}\Delta t}{\Delta z} \leq 1.
\end{align}

\noindent where the left hand side is called the Courant number $C$. While the step size of the solver internally is not always available or desirable to control, a reasonable approach is to strive for a step size $\Delta t$ that equals the sampling time (time resolution of interest). Since the flow velocity in a circular pipe is $v(t) = {\Dot{m}(t)}/{(\rho A)}$ for incompressible flow and $\Delta z = m / (\rho A)$ we can rewrite Equation \eqref{eq:CFL} as 
\begin{align}
    \frac{v_{sup}\Delta t}{\Delta z} = \frac{\Dot{m}_{sup}\Delta t}{m} = \frac{\Delta t}{\tau} \leq 1
    \Longleftrightarrow \Delta t \leq \tau
\end{align}

\noindent where $\tau$ is the transport delay through the pipe. The Courant number also relates to the numerical accuracy of the discretization. A truncation error analysis \cite{leveque_finite_2002} using a Taylor expansion gives that the local truncation error for a finite volume upwind discretization is approximately 
\begin{align}
    \label{eq:truncation_error}
    T \approx \frac{v\Delta z}{2}\left(1 - \frac{v\Delta t}{\Delta z}\right)\frac{d^2 x}{d z^2} + \mathcal{O}(\Delta x^3),
\end{align}
\noindent i.e., for $C=1$ the truncation error due to discretization vanishes. Notably, the expression for the truncation error is similar to the diffusion equation, and the phenomenon is hence called numerical (or artificial) diffusion. While there exist many methods to minimize numerical diffusion and choose an optimal time step, for practical applications in district energy simulation, it is in most cases sufficient to aim for a maximal Courant number for the simulation that provides a reasonable trade-off between accuracy and stability for a given time-resolution.

\subsection{The district energy grid as a graph}
\begin{figure}[ht!]
    \centering
    \begin{tikzpicture}[scale=0.4]
\path[fill=lightgray!35] (2.5,1.5) circle (1.5cm);
\path[fill=lightgray!35] (8.5,1.5) circle (1.5cm);
\path[fill=lightgray!35] (14.5,1.5) circle (1.5cm);
\path[fill=lightgray!35] (20.5,1.5) circle (1.5cm);
\path[fill=lightgray!35] (8.5,7.5) circle (1.5cm);

\draw[fill=white] (2,2) circle (0.5cm);
\draw[fill=white] (8,2) circle (0.5cm);
\draw[fill=white] (14,2) circle (0.5cm);
\draw[fill=white] (20,2) circle (0.5cm);
\draw[fill=white] (8,8) circle (0.5cm);

\filldraw[fill=lightgray, draw=black] (3,1) circle (0.5cm);
\draw[fill=lightgray, draw=black] (9,1) circle (0.5cm);
\draw[fill=lightgray, draw=black] (15,1) circle (0.5cm);
\draw[fill=lightgray, draw=black] (21,1) circle (0.5cm);
\draw[fill=lightgray, draw=black] (9,7) circle (0.5cm);

\draw[thick] (2.5,2) -- (7.5,2);
\draw[thick] (8.5,2) -- (13.5,2);
\draw[thick] (14.5,2) -- (19.5,2);
\draw[thick] (8.5,2) -- (13.5,2);
\draw[thick] (8,2.5) -- (8,7.5);

\draw[thick, color=gray] (3.5,1) -- (8.5,1);
\draw[thick, color=gray] (9.5,1) -- (14.5,1);
\draw[thick, color=gray] (15.5,1) -- (20.5,1);
\draw[thick, color=gray] (9.5,1) -- (14.5,1);
\draw[thick, color=gray] (9,1.5) -- (9,6.5);

\draw (13.5,3.5) rectangle (15.5,5.5);
\draw[thick, ->] (14,2.5) -- (14,3.5);
\draw[thick, <-, color=gray] (15,1.5) -- (15,3.5);

\draw (4.5,6.5) rectangle (6.5,8.5);
\draw[thick, <-] (6.5,8) -- (7.5,8);
\draw[thick, ->, color=gray] (6.5,7) -- (8.5,7);

\draw (19.5,3.5) rectangle (21.5,5.5);
\draw[thick, ->] (20,2.5) -- (20,3.5);
\draw[thick, <-, color=gray] (21,1.5) -- (21,3.5);

\draw[fill=white] (1.5,3.5) -- (1.5,5.0) -- (1.5,5.0) -- (2.0,4.5) -- (2.0,4.5) -- (2.0,5.0) -- (2.0,5.0) -- (2.5,4.5) -- (2.5,4.5) -- (2.5,5.0) -- (2.5,5.0) -- (3.0,4.5) -- (3.0,4.5) -- (3.0,6.0) -- (3.0,6.0) -- (3.5,6.0) -- (3.5,6.0) -- (3.5,3.5) -- (3.5,3.5) -- (1.5,3.5);
\draw[thick, ->] (2.0,3.5) -- (2.0,2.5) ;
\draw[thick, <-, color=gray] (3.0,3.5) -- (3.0,1.5);

\end{tikzpicture}
    \caption{Graphical representation of junctions, pipes, consumers, and one producer, where the shaded areas are referred to as nodes in the graph.}
    \label{fig:dh_graph_1}
\end{figure}
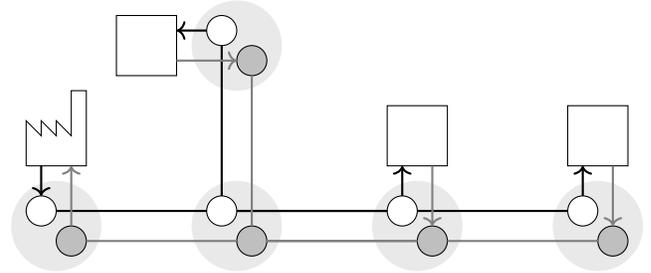

In Figure \ref{fig:dh_graph_1} a schematic view of a minimal district energy grid with three consumers and one producer is shown. The dashed circles represent nodes -- connection points for junctions, consumers, or producers. The lines between the nodes are called edges and represent pipes. This minimal grid will be used as an example in the theoretical part, whereas a city-scale grid will be used in the results section to show the method's applicability. 

\begin{figure}[ht!]
    \centering
    \begin{tikzpicture}[scale=0.4]

\draw[thick] (3.5,1.5) -- (7.5,1.5);
\node at (5.5,2.0) {$e_1$}; 
\draw[thick] (9.5,1.5) -- (13.5,1.5);
\node at (11.5,2.0) {$e_3$}; 
\draw[thick] (15.5,1.5) -- (19.5,1.5);
\node at (17.5,2.0) {$e_4$}; 
\draw[thick] (8.5,2.5) -- (8.5,6.5);
\node at (9.0,4.5) {$e_2$}; 

\draw (13.5,3.5) rectangle (15.5,5.5);
\draw[thick] (14.5,2.5) -- (14.5,3.5);

\draw (4.5,6.5) rectangle (6.5,8.5);
\draw[thick] (6.5,7.5) -- (7.5,7.5);

\draw (19.5,3.5) rectangle (21.5,5.5);
\draw[thick] (20.5,2.5) -- (20.5,3.5);

\draw[fill=white] (1.5,3.5) -- (1.5,5.0) -- (1.5,5.0) -- (2.0,4.5) -- (2.0,4.5) -- (2.0,5.0) -- (2.0,5.0) -- (2.5,4.5) -- (2.5,4.5) -- (2.5,5.0) -- (2.5,5.0) -- (3.0,4.5) -- (3.0,4.5) -- (3.0,6.0) -- (3.0,6.0) -- (3.5,6.0) -- (3.5,6.0) -- (3.5,3.5) -- (3.5,3.5) -- (1.5,3.5);
\draw[thick] (2.5,3.5) -- (2.5,2.5);

\draw[fill=lightgray] (2.5,1.5) circle (1.0cm);
\node at (2.5,1.5) {$n_1$}; 
\draw[fill=lightgray] (8.5,1.5) circle (1.0cm);
\node at (8.5,1.5) {$n_2$}; 
\draw[fill=lightgray] (14.5,1.5) circle (1.0cm);
\node at (14.5,1.5) {$n_4$}; 
\draw[fill=lightgray] (20.5,1.5) circle (1.0cm);
\node at (20.5,1.5) {$n_5$}; 
\draw[fill=lightgray] (8.5,7.5) circle (1.0cm);
\node at (8.5,7.5) {$n_3$};

\end{tikzpicture}
    \caption{Graph representation of a district energy grid with nodes weighted by $n_i$ and edges by $e_i$.}
    \label{fig:dh_graph_2}
\end{figure}
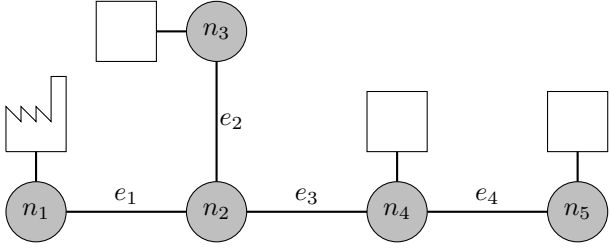

The flows are assumed as balanced so that the supply flow equals the return flow for each node. A simplified schematic view of the same grid can be seen in Figure \ref{fig:dh_graph_2}, where the labels $n_i$ and $e_i$ are added for nodes and edges. 
The grid can be represented by an incidence matrix, where the columns represent edges and rows represent nodes
\begin{align}
    \mathbf{M}_w(i,j) = 
    \begin{cases}
    \sqrt{w_{ij}}, \hspace{0.1cm} \text{if node $n_j$ is the target of $e_i$}\\
    -\sqrt{w_{ij}}, \hspace{0.1cm}  \text{if node $n_j$ is the source of $e_i$}\\
    0, \hspace{0.1cm}  \text{elsewhere}
    \end{cases}
\end{align}

Consumers and producers are not included in the incidence matrix in this paper. When the weights are chosen as $w_{ij} = 1$, the incidence matrix is unweighted. For the example grid, the unweighted incidence matrix is
\begin{align}
\mathbf{M} = 
\begin{blockarray}{ccccc}
 & e_1 & e_2 & e_2 & e_4 \\
\begin{block}{c[cccc]}
n_1 & -1 &  0 &  0 &  0\\
n_2 &  1 & -1 & -1 &  0\\
n_3 &  0 &  1 &  0 &  0\\
n_4 &  0 &  0 &  1 & -1\\
n_5 &  0 &  0 &  0 &  1\\
\end{block}
\end{blockarray}
\end{align}
The matrix $\mathbf{L}_w = \mathbf{M}_w \mathbf{M}_w^\top$ is called the weighted graph Laplacian, which is undirected, symmetric, and positive definite. The diagonal matrix with the diagonal elements from the weighted graph Laplacian is the degree matrix $\mathbf{D}_w$ and $\mathbf{W} = \mathbf{L}_w - \mathbf{D}_w$ is the weighted adjacency matrix. For unweighted matrices and vectors the subscript is dropped in the notation. 

For the model reduction performed in this paper, two different Laplacian matrices are used -- the flow Laplacian $\mathbf{L}_f = \mathbf{M}_f \mathbf{M}_f^\top$ where the weights correspond to the maximum flow rate through each pipe, and the mass Laplacian $\mathbf{L}_m = \mathbf{M}_m \mathbf{M}_m^\top$ where the weights correspond to the mass of water within each pipe. 

\subsection{Advection on a graph} 
The graph itself can be interpreted as a finite volume discretization, where the edges represent mass flow rate and the nodes represent the masses of the volumes. Advection on graphs for the general case is described in \cite{chapman_advection_2011}. The thermal dynamics of the supply and return pipes are modeled separately. To model advection, the modified incidence matrix 
\begin{align}
    \mathbf{M}_o(i,j) = 
    \begin{cases}
    1, \hspace{0.5cm} \mathbf{M}(i,j) = -1\\
    0, \hspace{0.5cm} \text{elsewhere}
    \end{cases}
\end{align}
and the flow matrix $\mathbf{U}_d = \text{diag}([u_1, \hdots, u_m])$ where $u_i$ is the flow on edge $e_i$, are introduced. $\mathbf{U}_{do}$ is the diagonal matrix of flows exiting a node from the supply to the return pipe or vice versa. $\mathbf{U}_{di}$ the diagonal matrix of flows entering each node and $\mathbf{x}_{in}$ the temperature of flows entering each node. Equation \eqref{eq:discretized_pipe} can then be written in matrix form as
\begin{align}
    \label{eq:state_space}
    \begin{split}
    \dot{\mathbf{x}}(t) = \mathbf{M}_d^{-1} (\mathbf{M} \mathbf{U}_d(t) \mathbf{M}_o^\top - \mathbf{U}_{do}(t)) \mathbf{x}(t)\\+ \mathbf{M}_d^{-1}\mathbf{U}_{di}(t) \mathbf{x}_{in}(t) 
    \end{split}
\end{align}
Equation \eqref{eq:state_space} can be recognized as a time-varying state space model, that can be written on the form
\begin{align}
    \begin{split}
    \dot{\mathbf{x}}(t) = \mathbf{A}(k)\mathbf{x}(t) + \mathbf{B}\mathbf{u}(t) \\
    \mathbf{y}(t) = \mathbf{C} \mathbf{x}(t)
    \end{split}
\end{align}
with
\begin{align}
    \begin{split}
    \mathbf{A}(t) = \mathbf{M}_d^{-1} (\mathbf{M} \mathbf{U}_d(t) \mathbf{M}_o^\top - \mathbf{U}_{do}(t))\\
    \mathbf{B} = \mathbf{M}_d^{-1}, \hspace{0.5cm} \mathbf{C} = \mathbf{I}\\
    \mathbf{u}(t) = \mathbf{u}_{in}(t) \circ \mathbf{x}_{in}(t) 
    \end{split}
\end{align}
Since there are sometimes long pipes between nodes (i.e., no connected consumers for a long stretch of piping) in district energy grids, the graph can be oversampled by introducing intermediate nodes for edges corresponding to long pipes.

\subsection{Low rank approximations}
A low rank approximation of the state space system means that the calculations can be performed in a lower dimensional space $\mathbb{R}^k$ with $k<n$ where $n$ is the number of states (nodes in the graph). The resulting number of edges from the reduction is $l$. The reduction matrices for a graph need to act on both the nodes and the edges by contracting nodes and removing edges within the contracted set, and thus two transformation matrices $\mathbf{P}_n, \mathbf{P}_e$, of sizes $k \times n, m \times l$ respectively, are needed. For the reduced state space model, using the notation from Equation \eqref{eq:state_space} and dropping the time index for more compact notation, the reduced versions of respective vectors and matrices become
\begin{align}
    \label{eq:transformations}
    \begin{split}
    \Tilde{\mathbf{M}}_d = \mathbf{P}_n \mathbf{M}_d \mathbf{P}_n^\top\\ 
    \Tilde{\mathbf{M}}= \mathbf{P}_n \mathbf{M} \mathbf{P}_e^\top, \hspace{0.2cm} 
    \Tilde{\mathbf{M}}_o = \mathbf{P}_n \mathbf{M}_o \mathbf{P}_e^\top \\
    \Tilde{\mathbf{U}}_d = \mathbf{P}_e \mathbf{U}_d \mathbf{P}_e^\top, \hspace{0.2cm}  \Tilde{\mathbf{U}}_{do} = \mathbf{P}_n \mathbf{U}_{do} \mathbf{P}_n^\top\\
    \Tilde{\mathbf{B}} = \mathbf{P}_n \mathbf{B} \mathbf{P}_n^\top, \hspace{0.2cm} \Tilde{\mathbf{u}} = \mathbf{P}_n \mathbf{u} \end{split}
\end{align}


Some special care needs to be taken for the reduced states $\Tilde{\mathbf{x}}$ with regards to energy conservation, since the temperature of the reduced state should be the average temperature of the nodes scaled by the mass of each node
\begin{align}
    \label{eq:x_transform}
    \Tilde{x}_i = \frac{\sum_{j \in C_i}{m_j x_j}}{\sum_{j \in C_i}{m_j}},
\end{align}
or in matrix form
\begin{align}
    \Tilde{\mathbf{x}} = (\mathbf{P}_n \mathbf{M}_d \mathbf{P}_n^\top)^{-1} \mathbf{P}_n  \mathbf{M}_d \mathbf{x}.
\end{align}

The reduced system can thus be written as 
\begin{align}
    \begin{split}
    \Tilde{\mathbf{x}}(t) = \Tilde{\mathbf{A}}(t)\Tilde{\mathbf{x}}(t) + \Tilde{\mathbf{B}}\Tilde{\mathbf{u}}(t)\\
    \mathbf{y}(t) = \mathbf{P}_n^\top \Tilde{\mathbf{x}}(t)
    \end{split}
\end{align}
where
\begin{align}
\Tilde{\mathbf{A}}(t) = \Tilde{\mathbf{M}}_d^{-1} (\Tilde{\mathbf{M}} \Tilde{\mathbf{U}}_d(t) \Tilde{\mathbf{M}}_o^\top - \Tilde{\mathbf{U}}_{do}(t))
\end{align}
Note that since $\mathbf{y}(t)$ is the lifted state vector in $\mathbb{R}^n$, there is a direct interpretation with regards to the states of the original graph.

\section{SPECTRAL GRAPH CLUSTERING}
Coarsening the spatial discretization of the grid corresponds to contracting nodes by adding the masses of the contracted nodes and removing the edges connecting contracted nodes. This can be seen as a graph clustering problem -- the original nodes of the graph should be assigned to the cluster $\mathcal{N}_i$ where the temperature is averaged within the cluster. The clustering of the reduced grid should minimize the truncation error for the advection equation with some stability margin with regards to the CFL condition, corresponding to choosing a target maximum Courant number $C_\tau$. 

The number of clusters needs to be approximated or chosen iteratively. When the number of clusters is decided, the minimization problems equals minimizing the sum of Courant numbers for the reduced problem
\begin{align}
    C_{\text{tot}} = \Delta t \sum_i^k{C_i} = \Delta t \sum_{i \leftrightarrow j}{\left(\frac{\Dot{m}_{ij}}{m_i} + \frac{\Dot{m}_{ij}}{m_j}\right)}
\end{align}
where $i \leftrightarrow j$ denotes that the clusters are connected and $\Delta t$ is the wanted step size.

\begin{figure}[ht!]
    \centering
    \begin{tikzpicture}[scale=0.4]


\draw[thick] (3.5,1.5) -- (7.5,1.5);
\node at (5.5,2.0) {$0.5$}; 
\draw[thick] (9.5,1.5) -- (13.5,1.5);
\node at (11.5,2.0) {$0.3$}; 
\draw[thick] (15.5,1.5) -- (19.5,1.5);
\node at (17.5,2.0) {$0.1$}; 
\draw[thick] (8.5,2.5) -- (8.5,6.5);
\node at (9.3,4.5) {$0.2$}; 

\draw[fill=white] (13.5,3.5) rectangle (15.5,5.5);
\draw[thick] (14.5,2.5) -- (14.5,3.5);

\draw[fill=white] (4.5,6.5) rectangle (6.5,8.5);
\draw[thick] (6.5,7.5) -- (7.5,7.5);

\draw[fill=white] (19.5,3.5) rectangle (21.5,5.5);
\draw[thick] (20.5,2.5) -- (20.5,3.5);

\draw[fill=white] (1.5,3.5) -- (1.5,5.0) -- (1.5,5.0) -- (2.0,4.5) -- (2.0,4.5) -- (2.0,5.0) -- (2.0,5.0) -- (2.5,4.5) -- (2.5,4.5) -- (2.5,5.0) -- (2.5,5.0) -- (3.0,4.5) -- (3.0,4.5) -- (3.0,6.0) -- (3.0,6.0) -- (3.5,6.0) -- (3.5,6.0) -- (3.5,3.5) -- (3.5,3.5) -- (1.5,3.5);
\draw[thick] (2.5,3.5) -- (2.5,2.5) ;

\draw[fill=lightgray] (2.5,1.5) circle (1.0cm);
\node at (2.5,1.5) {3}; 
\draw[fill=lightgray] (8.5,1.5) circle (1.0cm);
\node at (8.5,1.5) {3}; 
\draw[fill=lightgray] (14.5,1.5) circle (1.0cm);
\node at (14.5,1.5) {6}; 
\draw[fill=lightgray] (20.5,1.5) circle (1.0cm);
\node at (20.5,1.5) {1}; 
\draw[fill=lightgray] (8.5,7.5) circle (1.0cm);
\node at (8.5,7.5) {3};

\end{tikzpicture}
    \caption{Graph with edge and node weights.}
    \label{fig:mincut}
\end{figure}
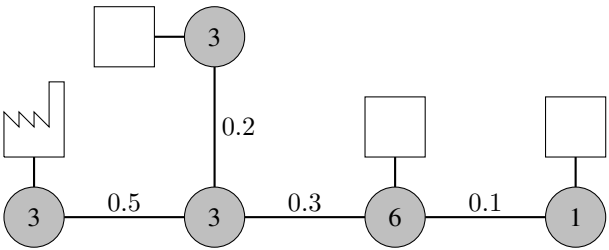

To deal with the problem, some additional notation is introduced. For a cut in a graph $\mathcal{G} = (\mathcal{N}, \mathcal{E})$ consisting of nodes $\mathcal{N}$ and edges $\mathcal{E}$, the weight of the cut is defined as the sum of the weights of the edges separating $\mathcal{N}_1, \mathcal{N}_2$ 
\begin{align}
    \text{cut}(\mathcal{N}_1, \mathcal{N}_2) = \sum_{i \in \mathcal{N}_1, j \in \mathcal{N}_2}{w_{ij}}
\end{align}
Nodes are weighted with $m_i$, and thus the total mass of a cluster is 
\begin{align}
    \text{mass}(\mathcal{N}_i) = \sum_{i \in \mathcal{N}_i}{m_i}.
\end{align}
To illustrate the clustering problem for advection on a graph, it is clarifying to look at the clustering of the graph to two clusters $\mathcal{N}_1, \mathcal{N}_2$ -- i.e. the cut is a bisection. Starting from the previously introduced graph of a district energy grid, edge weights corresponding to maximum flow rate through each pipe, and node weights corresponding to the mass of each node are added, as can be seen in Figure \ref{fig:mincut}.

\begin{figure}[ht!]
    \centering
    \begin{tikzpicture}[scale=0.4]

\path[fill=lightgray!35] (6.0,3.5) ellipse (5 and 6);
\path[fill=lightgray!35] (17.5,3.0) ellipse (5 and 4);
\node at (6.0,3.5) {9};
\node at (17.5,3.0) {7};

\draw[line width=5, dashed, gray] (10.5,-2) -- (13.0,8);

\draw[thick] (3.5,1.5) -- (7.5,1.5);
\node at (5.5,1.5) {$//$}; 
\draw[thick] (9.5,1.5) -- (13.5,1.5);
\node at (11.5,2.0) {$0.3$}; 
\draw[thick] (15.5,1.5) -- (19.5,1.5);
\node at (17.5,1.5) {$//$}; 
\draw[thick] (8.5,2.5) -- (8.5,6.5);
\node at (8.5,4.5) {$//$}; 

\draw[fill=white] (13.5,3.5) rectangle (15.5,5.5);
\draw[thick] (14.5,2.5) -- (14.5,3.5);

\draw[fill=white] (4.5,6.5) rectangle (6.5,8.5);
\draw[thick] (6.5,7.5) -- (7.5,7.5);

\draw[fill=white] (19.5,3.5) rectangle (21.5,5.5);
\draw[thick] (20.5,2.5) -- (20.5,3.5);

\draw[fill=white] (1.5,3.5) -- (1.5,5.0) -- (1.5,5.0) -- (2.0,4.5) -- (2.0,4.5) -- (2.0,5.0) -- (2.0,5.0) -- (2.5,4.5) -- (2.5,4.5) -- (2.5,5.0) -- (2.5,5.0) -- (3.0,4.5) -- (3.0,4.5) -- (3.0,6.0) -- (3.0,6.0) -- (3.5,6.0) -- (3.5,6.0) -- (3.5,3.5) -- (3.5,3.5) -- (1.5,3.5);
\draw[thick] (2.5,3.5) -- (2.5,2.5) ;

\draw[fill=lightgray] (2.5,1.5) circle (1.0cm);
\draw[fill=lightgray] (8.5,1.5) circle (1.0cm);
\draw[fill=lightgray] (14.5,1.5) circle (1.0cm);
\draw[fill=lightgray] (20.5,1.5) circle (1.0cm);
\draw[fill=lightgray] (8.5,7.5) circle (1.0cm);

\end{tikzpicture}
    \caption{A balanced cut, corresponding to minimization of the sum of Courant numbers, where the node masses are summed to a cluster mass.}
    \label{fig:contraction}
\end{figure}
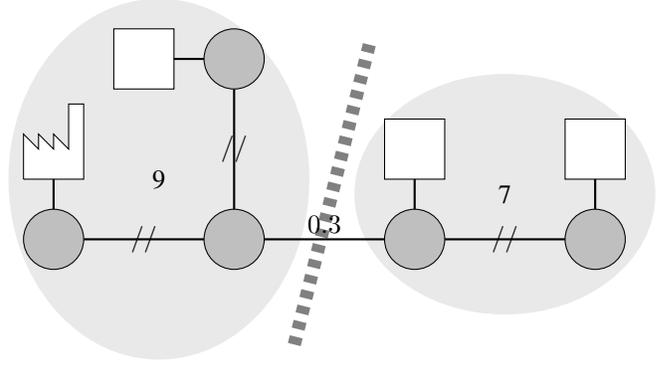

The minimum sum of Courant numbers is achieved when the cut is small while the masses of the two clusters are balanced. In the graph, this corresponds to cutting along the weight that is $0.3$ giving a sum of Courant numbers $C_1 + C_2 \approx 0.08\Delta t$, and is shown in Figure \ref{fig:contraction}.
With the introduced notation, this corresponds to a balanced cut problem minimizing the expression 
\begin{align}
    \mathcal{R} = \frac{ \text{cut}(\mathcal{N}_1, \mathcal{N}_2)}{ \text{mass}(\mathcal{N}_1) \text{mass}(\mathcal{N}_2)},
\end{align}
where $\Delta t$ is assumed constant and thus not relevant for the optimization. In matrix form, this can be expressed using the diagonal matrix $\mathbf{M}_d$ where $\mathbf{M}_{d,ii} = \frac{1}{2}\mathbf{L}_{m,ii}$ where $\mathbf{L}_m$ is the mass Laplacian, and $\mathbf{L}_f$ is the flow Laplacian. 
\begin{align}
    \label{eq:optimization}
    \begin{tabular}{c c}
        \text{minimize} & $\mathbf{v}^\top \mathbf{L}_f \mathbf{v}$\\
        \text{s.t} & $\mathbf{v}^\top \mathbf{M}_d \mathbf{v} = m(\mathcal{G})$\\
         & $\mathbf{1} \mathbf{M}_d \mathbf{v} = 0$.\\
    \end{tabular}
\end{align}
Following the procedure in \cite{shi_normalized_2000}, the minimization can be expressed in a more compact form as 
\begin{align}
    \label{eq:rayleigh}
    \underset{\mathbf{v}}{\text{minimize }} 
    \frac{\mathbf{v}^\top \mathbf{L}_f \mathbf{v}}{\mathbf{v}^\top \mathbf{M}_d \mathbf{v}} = \frac{\sum_{}\Dot{m}_{ij}(v_i - v_j)^2}{\sum m_i v_i^2}.
\end{align}
That the expression minimizes the Courant number can be shown for $k=2$ by introducing the indicator vector $\mathbf{v}$  where $v_{i}=1$ if the corresponding node $n_i \in \mathcal{N}_1$ and $v_{i}=-1$ if $n_i \in \mathcal{N}_2$, in the expression above. The right hand side of Equation \eqref{eq:rayleigh} can then be rewritten as
\begin{align}
    \frac{\sum_{}\Dot{m}_{ij}(v_i - v_j)^2}{\sum m_i v_i^2} = 4\frac{\Dot{m}_{12}}{m_{\mathcal{N}1}} + 4\frac{\Dot{m}_{21}}{m_{\mathcal{N}2}} = \frac{4}{\Delta t}(C_1 + C_2), 
\end{align}
i.e. the sum of Courant numbers scaled by a constant. This combinatorial optimization problem is known to be NP-hard \cite{von_luxburg_tutorial_2007}, but the optimum can be approximated by relaxing the problem so that $\mathbf{v}$ contains real numbers, and the partitions are chosen from the sign of $\mathbf{v}$. The expression on the left hand side of Equation \eqref{eq:rayleigh} is the generalized Rayleigh quotient and is minimized by the smallest eigenvector of the generalized eigenvalue problem $\mathbf{L}_f\mathbf{v} = \lambda\mathbf{M_d}\mathbf{v}$. The first eigenvector is trivially $\mathbf{v}_1 = \mathbf{1}/\sqrt{n}$, but the second eigenvector $\mathbf{v}_2$ called the Fiedler vector, that is orthogonal to $\mathbf{v}_1$, solves the minimization problem.

Following the procedure in \cite{von_luxburg_tutorial_2007}, the method can be extended to $k$ clusters by including the first $k$ vectors of the generalized eigenvalue problem, forming the $n \times k$ matrix of eigenvectors $\mathbf{V}_k = [\mathbf{v}_1 \hdots \mathbf{v}_k]$ and performing k-means clustering on the rows -- i.e. viewing the nodes as points in the $k$-dimensional space spanned by the rows of $\mathbf{V}_k$. The maximal Courant number of the reduced-order system can be calculated as
\begin{align}
\Tilde{C}_{\text{max}} =
\text{max}(\text{diag}(0.5 \Delta t \Tilde{\mathbf{M}}_d^{-1}\Tilde{\mathbf{L}}_f)),
\end{align}
where the number of clusters should be adjusted so that $\Tilde{C}_{\text{max}} \approx C_\tau$. The clustering process is summarized in Algorithm \ref{alg:clustering}.
\begin{algorithm}[ht!]
\caption{Spectral graph clustering of district energy network}
\label{alg:clustering}
$\bullet$ $k \gets$ {user specified number of clusters}\\
$\bullet$ Calculate graph Laplacians $ \mathbf{L}_f, \mathbf{L}_m$\\
$\bullet$ $\mathbf{M}_d \gets \text{Diagonal}(\text{diag}(\mathbf{L}_m))$\\
$\bullet$ Solve the generalized eigenvalue problem  $\mathbf{L}_f \mathbf{v} = \lambda \mathbf{M}_d \mathbf{v}$\\
$\bullet$ $\mathbf{V}_k \gets [\mathbf{v}_1, \hdots, \mathbf{v}_k]$\\
$\bullet$ Let $\mathbf{y}_i = \mathbf{V}_k(i,:)$\\
$\bullet$ Assign each $\mathbf{y}_i$ to cluster $\mathcal{N}_j$ by performing a k-means clustering on the rows of $\mathbf{V}_k$\\
$\bullet$ Adjust the number of clusters if $C_{\text{max}} \not\approx C_\tau$
\end{algorithm}

The reduction matrices $\mathbf{P}_n \in \mathbb{R}^{k \times n}, \mathbf{P}_e \in \mathbb{R}^{m \times l}$ calculated from a given clustering serve two separate purposes, namely 1) merging the nodes of each cluster, and 2) removing the edges within each cluster, where
\begin{align}
    \mathbf{P}_n(i,j) = 
    \begin{cases}
    1, \hspace{0.5cm} n_j \in \mathcal{N}_i\\
    0, \hspace{0.5cm} \text{elsewhere}\\
    \end{cases}
\end{align}
and
\begin{align}
    \mathbf{P}_e(i,j) = 
    \begin{cases}
    1, \hspace{0.5cm} e_i \in \mathcal{N}_j \rightarrow \mathcal{N}_{k \neq j} \\
    0, \hspace{0.5cm} \text{elsewhere}
    \end{cases}
\end{align}
The reduced-order system matrices in Equation \eqref{eq:transformations} can then be computed.

\subsection{A note on numerical methods}
The computational bottlenecks for large grids in Algorithm \ref{alg:clustering} are the calculation of the eigenvalues and the memory requirements of the matrices and corresponding algorithms. However, only the first $k$ eigenvalues and corresponding eigenvectors are needed, and the graph Laplacian matrix is sparse. Storing the graph Laplacian matrices in a sparse format and using iterative algorithms such as Lanczos algorithm allows for efficient calculations of the first $k$ eigenvalues and corresponding eigenvectors when the number of clusters is sufficiently small. For reference, on a standard laptop with an 8th generation i7 Intel processor and 16GB of RAM, some elapsed CPU times for the clustering algorithm are presented in Table \ref{tab:computational}. 
\begin{table}[ht!]
    \centering
    \caption{Elapsed CPU time for the clustering algorithm}
    \begin{tabular}{|c|c|c|}
         \hline
         Number of nodes $n$ & Number of clusters $k$ & CPU time\\
         \hline
         3180 & 50 & 0.7s\\
          & 150 & 1.8s\\
          & 500 & 26.5s\\
         \hline
    \end{tabular}
    
    \label{tab:computational}
\end{table}

\section{RESULTS}
To evaluate the clustering performance, a partial model of the district energy grid of Luleå, Sweden, was compared with the same model where the model order reduction provided in the paper was applied. The model comparison only includes the thermal dynamics of the supply piping. The advection equation and flows are similar on the return side, with an opposite flow direction, but the return temperatures of the consumers need to be modeled to get realistic results for the return piping, which is beyond the scope of the paper.  

The models were automatically generated from grid data, where the spatial discretization of the reference model corresponds to that each pipe was discretized with a uniform segment length $\Delta z$ when the pipe length exceeds $\Delta z$, elsewhere the pipe length provided by data was used. The original model consisted of about 3700 pipe sections and 1500 consumers. After some initial preprocessing, where nodes connected by pipes shorter than 10m were merged, the resulting model included 2715 pipes. 

Both models were generated as Julia code and run using BDF solvers provided by the package DifferentialEquations.jl \cite{rackauckas_differentialequationsjl_2017}. In this case, and the general case, there was no measured ground truth available -- the reference instead corresponded to what would be the result of directly modeling the grid from available geographical data with pipe models that are uniformly discretized using finite volumes. 

\begin{figure}[ht!]
    \centering
    \includegraphics[width=0.45\textwidth]{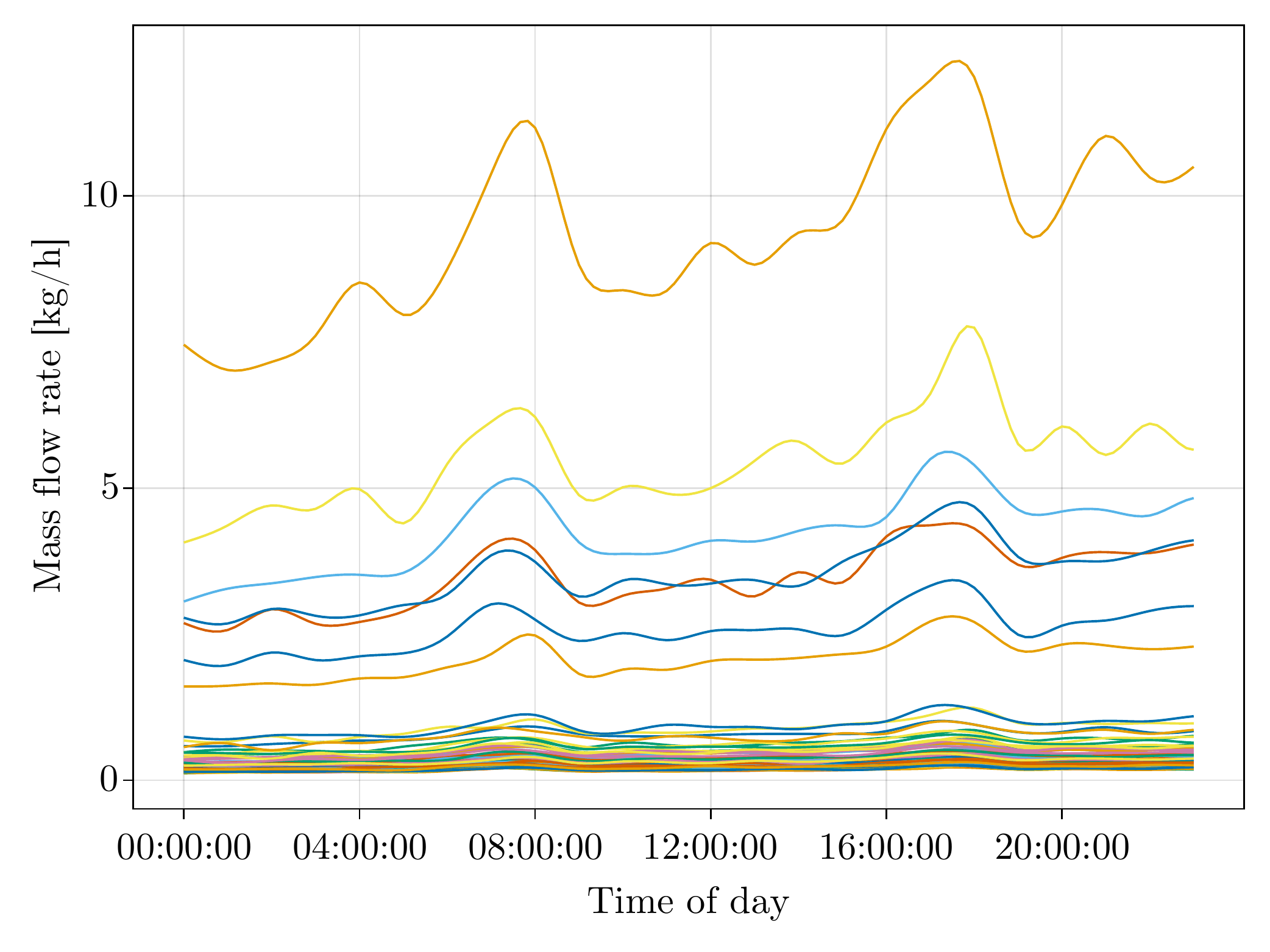}
    \caption{The flow demand of $100$ of the total of $1513$ consumers for the simulation time of $24h$.}
    \label{fig:consumers}
\end{figure}

The simulation scenario used was an increased outdoor temperature, with an increased demand of the consumers and a linearly increasing supply temperature. The demands of the consumers included morning and afternoon peaks, with random perturbations so that the demand pattern resembled what could be expected from real-life data. An example of the demand for $100$ different consumers is seen in Figure \ref{fig:consumers}.
The simulation time was set to $24h$, where the reference model used a time-step of $5s$ and the reduced model a time-step of $600s$. While the model was optimized with respect to the time-step, this was not necessarily the internal time step of the solver. The reduced model's speedup factor was around  90, resulting in the reduced-order model running at about 4000 times the real-time speed on a standard laptop computer. To evaluate how close the reduced model tracked the reference model, the states (temperatures) $\mathbf{x}$ of the original model are compared with the lifted states $\mathbf{P}_n^\top \Tilde{\mathbf{x}}$, at each successful time step of the larger step size, using the relative root mean square error (rRMSE)  
\begin{align}
    \text{rRMSE} = 100 \frac{1}{\Bar{x}}\sqrt{||(\mathbf{P}_n^\top \mathbf{\Tilde{x}} - \mathbf{x})||_2}.
\end{align}

\begin{figure}[ht!]
    \centering
    \includegraphics[width=0.45\textwidth]{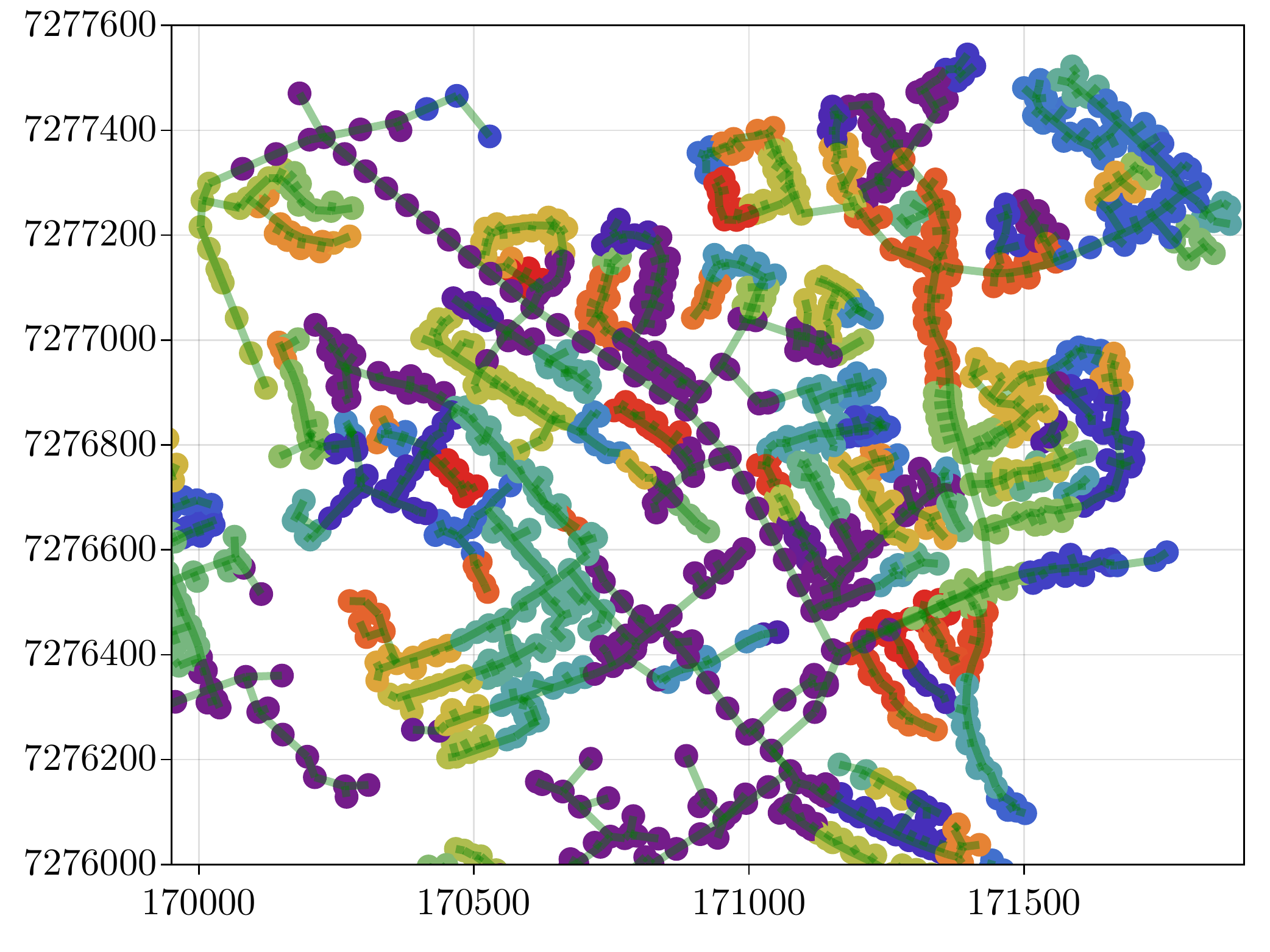}
    \caption{Example of clusters found by the spectral clustering algorithm.}
    \label{fig:clustering_result}
\end{figure}

The reduced-order model was generated using the algorithm provided in Algorithm \ref{alg:clustering} and the procedure described in the paper. An example of the clustering found by the algorithm for $k=150$ clusters is seen in Figure \ref{fig:clustering_result}. The number of clusters was chosen so that $C_{\text{max}} \approx 0.9$ for the time step size and maximum flow rate, corresponding to about $150$ clusters for the reduced model. 

\begin{figure}[ht!]
    \centering
    \includegraphics[width=0.5\textwidth]{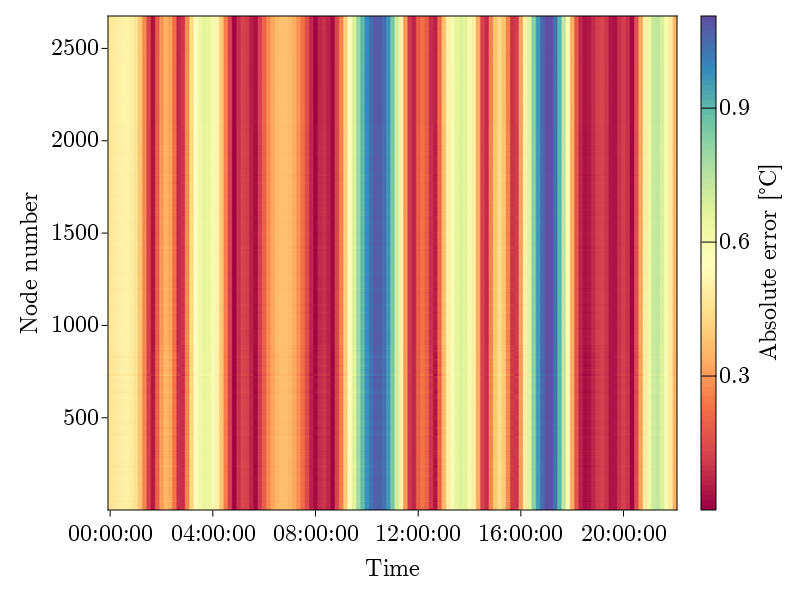}
    \caption{Heat map of the absolute error for each node, with node number on the $y$ axis, and time on the $x$ axis.}
    \label{fig:error}
\end{figure}

The first comparison used idealized pipe models only including the advection equation for thermal dynamics so that e.g., (non-numerical) diffusion and thermal losses were neglected. The simulation comparison resulted in a rRMSE of  $\approx 0.7\%$, corresponding to an absolute error within $1^\circ C$. A heat map of the absolute error can be seen in Figure \ref{fig:error}. 

From the truncation error analysis, it can be deduced that the reduced-order model does not necessarily have a larger truncation error so that the reduced model might be better in this regard. However, the dynamics of the more fine-grained model between the $600s$ time steps is lost in the comparison, that might be of interest for specific use cases. 

\begin{figure}[ht!]
    \centering
    \includegraphics[width=0.5\textwidth]{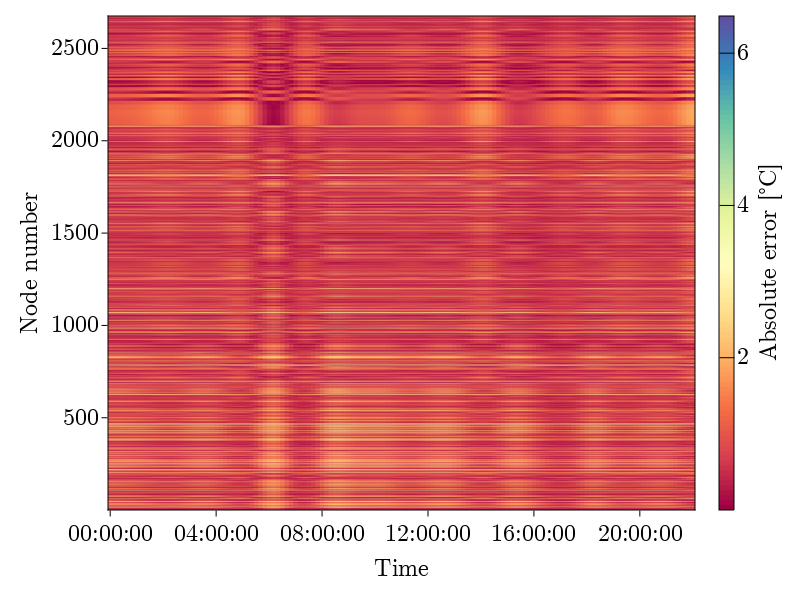}
    \caption{Heat map of the absolute error for each node, with node number on the $y$ axis, and time on the $x$ axis.}
    \label{fig:error_heat}
\end{figure}

In the second comparison shown in Figure \ref{fig:error_heat} a more realistic model with heat loss to the outside was used, otherwise using the same setup as in the preceding comparison. The resulting rRMSE was $2.3\%$ for this case, and as expected, some individual nodes -- typically for low consumption consumers -- were relatively far off due to averaging the temperature over the cluster.

In Table \ref{tab:computational_sim} a comparison between the elapsed CPU time and the rRMSE for some different choices for the number of clusters for 24h of simulation time, compared at the instances of $600s$ time steps is shown.

\begin{table}[ht!]
    \centering
    \caption{Elapsed CPU time and relative RMSE for 24h of simulation}
    \begin{tabular}{|c|c|c|c|}
         \hline
         Clusters $k$ & Step size & CPU time & rRMSE\\
         \hline
         No clustering & 5s & 2042s& \\
         500 & 600s & 508.9s& 2.9\%\\
         150 & 600s & 23.1s& 2.3\%\\
         50 & 600s & 5.9s& 3.0\%\\
         \hline
    \end{tabular}
    
    \label{tab:computational_sim}
\end{table}

Notably, more clusters did not necessarily result in a better rRMSE for a given time step, consistent with the theoretical truncation error.
\section{SUMMARY AND CONCLUSIONS}
This paper presents a novel method for reduced-order modeling of a district energy grid's thermal dynamics. The method preserves the total energy of the water in the system, and the states can be directly interpreted with the states of the original system. It is shown that the reduced model can run with a speedup factor of $\approx$ 90 with a corresponding rRMSE of 0.7\% for all nodes in the non-reduced grid for the advection only, and 2.3\% when including heat losses to surroundings, compared at the instances of the longer time step. 

While it is a promising result for district energy simulation, some theoretical background, bounds, and choices of parameters such as the number of clusters are in large parts omitted due to the format of the article and could be expanded on in the future. 

Moreover, the pressure and flow calculations are only briefly mentioned in the article but are of considerable interest for many use cases. While these calculations can be assumed to be static for the time resolutions assumed in this paper, e.g., pump characteristics, thermal storages, and consumer behavior needs to be included for a complete model -- a challenging problem in its own right.

\addtolength{\textheight}{-12cm}   


\section*{ACKNOWLEDGMENT}
The authors want to thank Luleå Energi AB and especially Fredrik Udén for discussions and making data available.

\printbibliography

@book{leveque_finite_2002,
	address = {Cambridge},
	series = {Cambridge {Texts} in {Applied} {Mathematics}},
	title = {Finite {Volume} {Methods} for {Hyperbolic} {Problems}},
	isbn = {978-0-521-00924-9},
	publisher = {Cambridge University Press},
	author = {LeVeque, Randall J.},
	year = {2002},
	doi = {10.1017/CBO9780511791253},
}

@article{falay_enabling_2020,
	title = {Enabling large-scale dynamic simulations and reducing model complexity of district heating and cooling systems by aggregation},
	volume = {209},
	issn = {0360-5442},
	doi = {10.1016/j.energy.2020.118410},
	language = {en},
	journal = {Energy},
	author = {Falay, Basak and Schweiger, Gerald and O’Donovan, Keith and Leusbrock, Ingo},
	month = oct,
	year = {2020},
	pages = {118410},
}

@inproceedings{chapman_advection_2011,
	title = {Advection on graphs},
	doi = {10.1109/CDC.2011.6161471},
	booktitle = {2011 50th {IEEE} {Conference} on {Decision} and {Control} and {European} {Control} {Conference}},
	author = {Chapman, Airlie and Mesbahi, Mehran},
	month = dec,
	year = {2011},
	note = {ISSN: 0743-1546},
	pages = {1461--1466},
}

@article{sartor_comparative_2018,
	title = {A comparative study for simulating heat transport in large district heating networks},
	volume = {36},
	issn = {03928764},
	doi = {10.18280/ijht.360140},
	number = {1},
	journal = {International Journal of Heat and Technology},
	author = {Sartor, Kevin and Thomas, David and Dewallef, P},
	month = mar,
	year = {2018},
	pages = {301--308},
}

@article{schweiger_district_2018,
	title = {District energy systems: {Modelling} paradigms and general-purpose tools},
	volume = {164},
	issn = {0360-5442},
	shorttitle = {District energy systems},
	doi = {10.1016/j.energy.2018.08.193},
	language = {en},
	journal = {Energy},
	author = {Schweiger, Gerald and Heimrath, Richard and Falay, Basak and O'Donovan, Keith and Nageler, Peter and Pertschy, Reinhard and Engel, Georg and Streicher, Wolfgang and Leusbrock, Ingo},
	month = dec,
	year = {2018},
	pages = {1326--1340},
}

@article{rackauckas_differentialequationsjl_2017,
	title = {{DifferentialEquations}.jl – {A} {Performant} and {Feature}-{Rich} {Ecosystem} for {Solving} {Differential} {Equations} in {Julia}},
	volume = {5},
	copyright = {Authors who publish with this journal agree to the following terms:    Authors retain copyright and grant the journal right of first publication with the work simultaneously licensed under a  Creative Commons Attribution License  that allows others to share the work with an acknowledgement of the work's authorship and initial publication in this journal.  Authors are able to enter into separate, additional contractual arrangements for the non-exclusive distribution of the journal's published version of the work (e.g., post it to an institutional repository or publish it in a book), with an acknowledgement of its initial publication in this journal.  Authors are permitted and encouraged to post their work online (e.g., in institutional repositories or on their website) prior to and during the submission process, as it can lead to productive exchanges, as well as earlier and greater citation of published work (See  The Effect of Open Access ).  All third-party images reproduced on this journal are shared under Educational Fair Use. For more information on  Educational Fair Use , please see  this useful checklist prepared by Columbia University Libraries .   All copyright  of third-party content posted here for research purposes belongs to its original owners.  Unless otherwise stated all references to characters and comic art presented on this journal are ©, ® or ™ of their respective owners. No challenge to any owner’s rights is intended or should be inferred.},
	issn = {2049-9647},
	doi = {10.5334/jors.151},
	language = {en},
	number = {1},
	journal = {Journal of Open Research Software},
	author = {Rackauckas, Christopher and Nie, Qing},
	month = may,
	year = {2017},
	note = {Number: 1
Publisher: Ubiquity Press},
	pages = {15},
}

@article{von_luxburg_tutorial_2007,
	title = {A tutorial on spectral clustering},
	volume = {17},
	issn = {0960-3174, 1573-1375},
	doi = {10.1007/s11222-007-9033-z},
	language = {en},
	number = {4},
	journal = {Statistics and Computing},
	author = {von Luxburg, Ulrike},
	month = dec,
	year = {2007},
	pages = {395--416},
}

@article{loukas_graph_2019,
	title = {Graph {Reduction} with {Spectral} and {Cut} {Guarantees}},
	volume = {20},
	issn = {1533-7928},
	number = {116},
	journal = {Journal of Machine Learning Research},
	author = {Loukas, Andreas},
	year = {2019},
	pages = {1--42},
}

@article{van_der_heijde_dynamic_2017,
	title = {Dynamic equation-based thermo-hydraulic pipe model for district heating and cooling systems},
	volume = {151},
	issn = {0196-8904},
	doi = {10.1016/j.enconman.2017.08.072},
	language = {en},
	journal = {Energy Conversion and Management},
	author = {van der Heijde, B. and Fuchs, M. and Ribas Tugores, C. and Schweiger, G. and Sartor, K. and Basciotti, D. and Müller, D. and Nytsch-Geusen, C. and Wetter, M. and Helsen, L.},
	month = nov,
	year = {2017},
	pages = {158--169},
}

@book{antoulas_approximation_2005,
	series = {Advances in {Design} and {Control}},
	title = {Approximation of {Large}-{Scale} {Dynamical} {Systems}},
	isbn = {978-0-89871-529-3},
	publisher = {Society for Industrial and Applied Mathematics},
	author = {Antoulas, Athanasios C.},
	month = jan,
	year = {2005},
	doi = {10.1137/1.9780898718713},
}

@article{todini_unified_2013,
	title = {Unified {Framework} for {Deriving} {Simultaneous} {Equation} {Algorithms} for {Water} {Distribution} {Networks}},
	volume = {139},
	issn = {0733-9429, 1943-7900},
	doi = {10.1061/(ASCE)HY.1943-7900.0000703},
	language = {en},
	number = {5},
	journal = {Journal of Hydraulic Engineering},
	author = {Todini, Ezio and Rossman, Lewis A.},
	month = may,
	year = {2013},
	pages = {511--526},
}

@article{Vandermeulen2018,
	title = {Controlling district heating and cooling networks to unlock flexibility: {A} review},
	volume = {151},
	issn = {03605442},
	doi = {10.1016/j.energy.2018.03.034},
	journal = {Energy},
	author = {Vandermeulen, Annelies and van der Heijde, Bram and Helsen, Lieve},
	year = {2018},
	note = {Publisher: Elsevier Ltd},
	pages = {103--115},
}

@book{trottenberg_multigrid_2000,
	address = {Amsterdam Heidelberg},
	edition = {1st edition},
	title = {Multigrid},
	isbn = {978-0-12-701070-0},
	language = {English},
	publisher = {Academic Press},
	author = {Trottenberg, Ulrich and Oosterlee, Cornelius W. and Schuller, Anton},
	month = dec,
	year = {2000},
}

@article{shi_normalized_2000,
	title = {Normalized {Cuts} and {Image} {Segmentation}},
	volume = {22},
	language = {en},
	number = {8},
	journal = {IEEE TRANSACTIONS ON PATTERN ANALYSIS AND MACHINE INTELLIGENCE},
	author = {Shi, Jianbo and Malik, Jitendra},
	year = {2000},
	pages = {18},
}

@article{simonsson_experiences_2021,
	title = {Experiences from {City}-{Scale} {Simulation} of {Thermal} {Grids}},
	volume = {10},
	copyright = {http://creativecommons.org/licenses/by/3.0/},
	url = {https://www.mdpi.com/2079-9276/10/2/10},
	doi = {10.3390/resources10020010},
	language = {en},
	number = {2},
	urldate = {2021-01-25},
	journal = {Resources},
	author = {Simonsson, Johan and Atta, Khalid Tourkey and Schweiger, Gerald and Birk, Wolfgang},
	month = feb,
	year = {2021},
	note = {Number: 2
Publisher: Multidisciplinary Digital Publishing Institute},
	pages = {10},
}

\end{document}